\documentclass[12pt]{article}

\usepackage[
backend=biber,
style=numeric-comp,
sorting=none]{biblatex}
\usepackage{caption}
\usepackage{subcaption}
\usepackage{graphicx}
\graphicspath{{figures/}}

\usepackage{amssymb}
\usepackage{amsmath} % Required for some math elements 

\usepackage{algorithmic}
\usepackage{array}
\usepackage{lipsum}
\usepackage[hidelinks]{hyperref}
\begin{document}

\title{Converting OpenStreetMap Data to Road Networks for Downstream Applications} % Title
%\author{Konstantin Akhmadeev, That Guy, Yet Another Guy (in alphabetic order)}
\author{Md Kaisar Ahmed}
\date{November 14, 2022} % Date for the report
\maketitle % Inserts the title, author and date

\begin{abstract}
%% Text of the abstract
We study how to convert OpenStreetMap data to road networks for downstream applications.
OpenStreetMap data has different formats. Extensible Markup Language (XML) is one of them. OSM data consist of nodes, ways, and relations. We process OSM XML data to extract the information of nodes and ways to obtain the map of streets of the Memphis area. 
% We parse the OSM data in such a way that gives us the whole map of the Memphis area. 
We can use this map for different downstream applications. 
% The steps that are included in this work downloading the Memphis area OSM data, understanding and parsing the OSM XML file, converting the nodes and ways information into the Pandas DataFrame, and visualizing these data into the whole map by using python's available data visualization libraries.
\end{abstract}

\section{Introduction}

OpenStreetMap (OSM)~\cite{OpenStreetMap} is referred to as the Wikipedia of the mapping of the world. People from all over the world contribute to build it. 
OSM 
% describes the ideal concept of every location we want to visit and 
represents the visual representation of the world. In this work, we detail how to convert OSM data to road networks that many downstream applications depend on.

\begin{figure}[hbt!]
    \centering\includegraphics[width=0.85\linewidth]{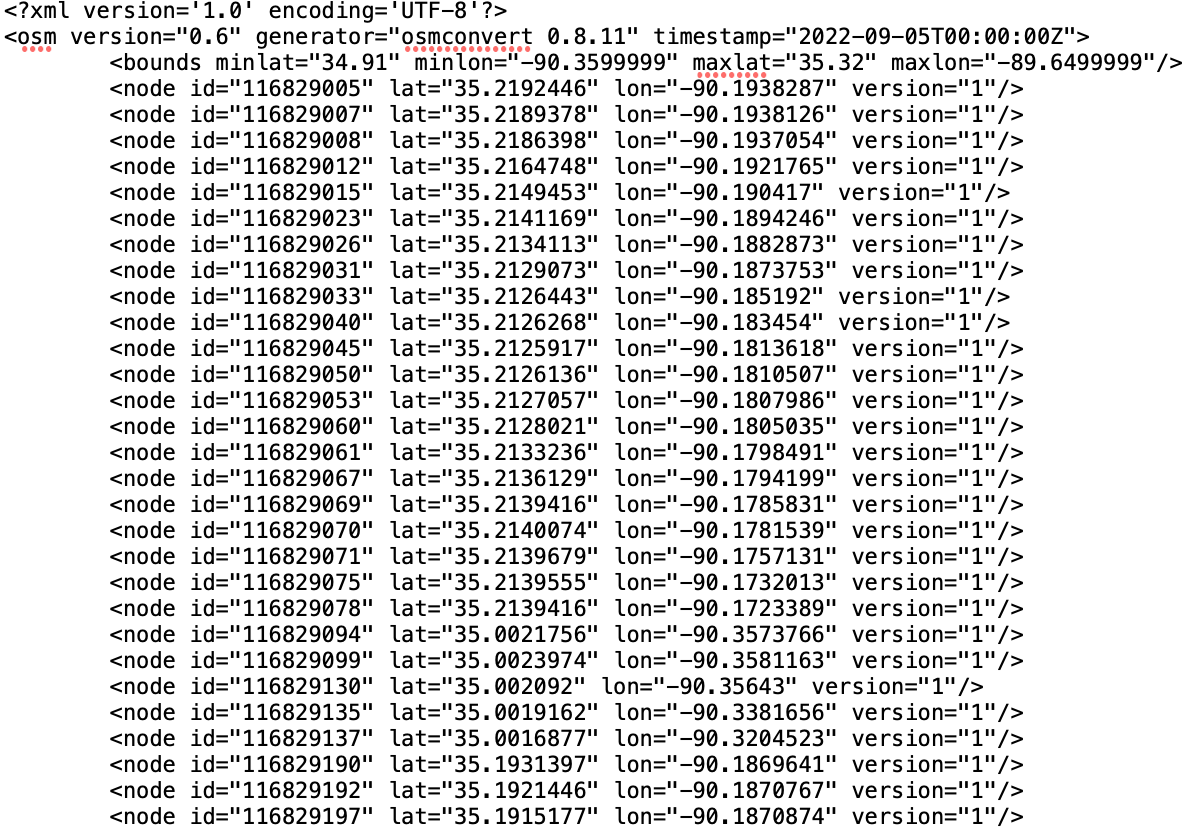}
    \caption{An example of raw OSM data in XML format.}
    \label{fig:XML file}
\end{figure}

\begin{figure}[hbt!]
    \centering\includegraphics[width=0.85\linewidth]{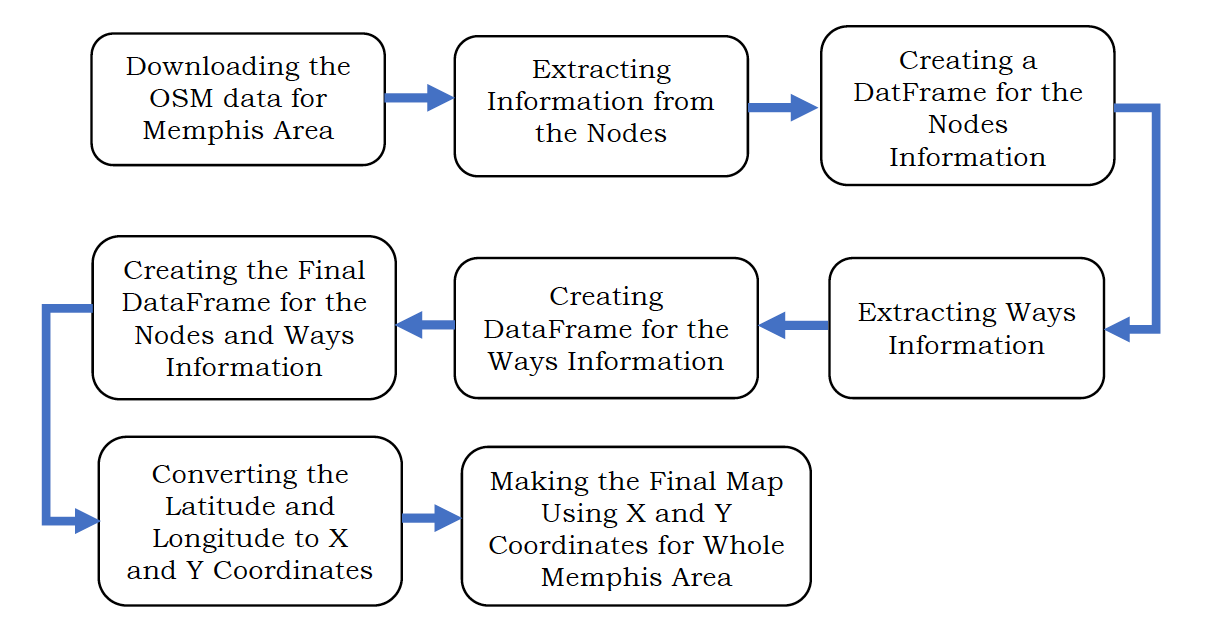}
    \caption{Flowchart of this project (using the Memphis area as an example).}
    \label{fig:flochart}
\end{figure}

OSM data has many formats, among which XML is a popular choice. 
OSM data is the collection of nodes, ways, and relations. 
Nodes contain the location in the WGS84 (World Geodetic Coordinate System 1984)\footnote{\url{https://wiki.openstreetmap.org/wiki/Converting_to_WGS84}} coordinate system specified by a pair of latitude and longitude. 
way is a collection of nodes (at least two). 
If we plot all nodes of a way by considering the latitude as the X axis and longitude as the Y axis, a line will be drawn.
If the first node and the last node of a way coincide, this way is considered closed. 
% An element is an ordered list of one or more nodes and ways. It indicates the link between a lake and its island or a couple of roads for a bus route. 
Various attributes can be assigned to each node, way, and relation. The attributes are represented as tags in OSM XML. Each tag is composed of a key and a value.
In Fig.~\ref{fig:XML file}, we show an example of the raw OSM data in the XML format. 
% We can further parse the XML file by using Python for getting the actual region of interest. 
The flowchart of this project is shown in Fig.~\ref{fig:flochart}.
The data of the Memphis area is downloaded from \url{https://extract.bbbike.org/}, which packs city-level OSM data.

\section{Nodes and Ways} 
\label{download}
A node represents a particular location on the surface of the earth that is identified by its latitude and longitude. 
At a minimum, a node has an ID number and a pair of latitude and longitude. 
The shape of a way can then be defined using nodes. 
The majority of the nodes do not have tags when they are used as intermediate points along ways. 
% Traffic signs are denoted, for instance, by highway=traffic signals. 
A way is an ordered list of at least two nodes that together define a polyline. 
% The representation of linear features like rivers and highways is done using ways. 

% \subsection{Downloading the OSM Data of Memphis}
% There are some required information including the format of the data, which is the area want to extract, and the email address need to be put on those fields in the website to download the data. Once the blanks are filled in, the authority of this website will send a mail with the link to the data set for the desired area. After downloading the data anyone can use the data for working on different applications.

\subsection{Extracting Node Information}
\begin{figure}[hbt!]
    \centering\includegraphics[width=.6\linewidth]{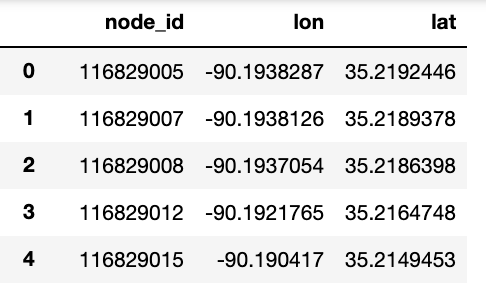}
    \caption{The first four nodes' information of the OSM data of the Memphis area.}
    \label{fig:node}
\end{figure}

To extract the information of nodes, we use the xml.etree.cElementTree package. This is one of the most popular python packages for parsing XML data. We know that a node consists of id, latitude, and longitude. With the help of the xml.etree.cElementTree package, we parse the node's information and convert it into the pandas DataFrame. 
% After converting the node's information into the pandas DataFrame we got three columns of the DataFrame. The name of the columns is node id, latitude, and longitude. 
An example is shown in Fig.~\ref{fig:node}.

\subsection{Extracting Way Information}
\begin{figure}[hbt!]
    \centering\includegraphics[width=.6\linewidth]{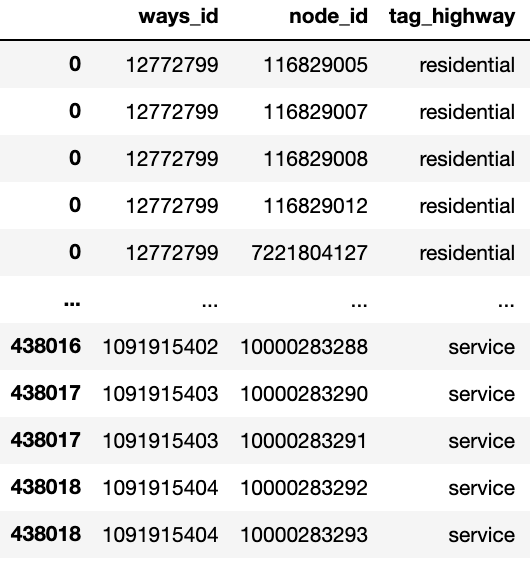}
    \caption{An example of the way information of the OSM data of the Memphis area.}
    \label{fig:ways}
\end{figure}

To extract the information of ways, again, we use xml.etree.cElementTree. 
A way contains id, the references of its constituting nodes references, and tags. 
Each tag has a key and a value. 
In this task, we extract only the highway keys: residential, service, tertiary, track, secondary, primary, tertiary link, secondary link, motorway link, primary link, motorway, trunk link, trunk, footway, construction, pedestrian, proposed, path, raceway, cycleway, living street, steps, abandoned, rest area, corridor, and platform. 
We put this information into pandas DataFrame for later use. 
An example is shown in Fig.~\ref{fig:ways}. 
% The columns for the ways in pandas DataFrame are ways id, nodes references, and the values of the tag highways.

\subsection{Merging Node and Way Information}
\begin{figure}[hbt!]
    \centering\includegraphics[width=.9\linewidth]{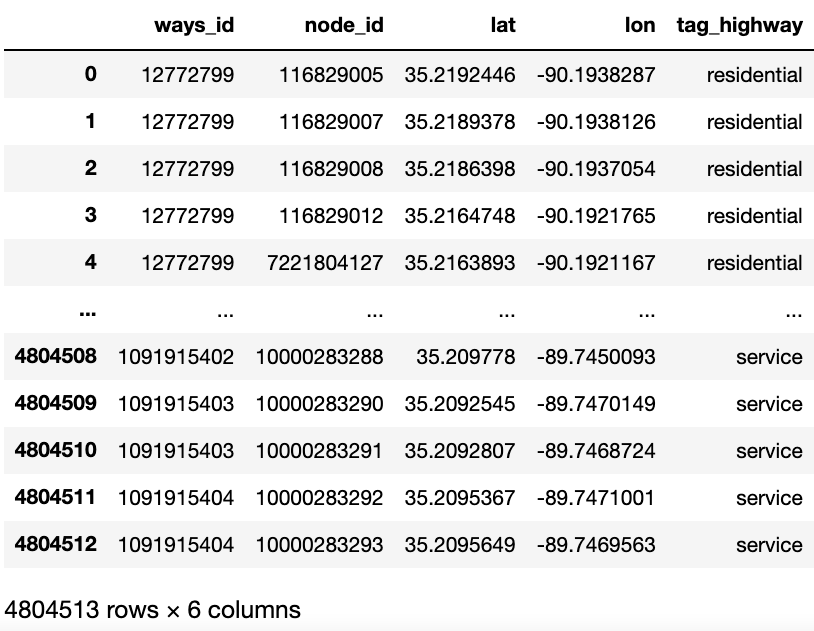}
    \caption{The final DataFrame for Combining the node and way information for the OSM Data of the Memphis area.}
    \label{fig:final}
\end{figure}

Next, we need to merge the node's information and way's information to get the final DataFrame. 
We know that single ways have multiple node references and we want to extract the latitude and longitude for those nodes' references. 
Once we are able to do that, we can plot the latitudes and longitudes according to the ways. 
Fig.~\ref{fig:final} shows the final DataFrame for putting the node and way information together.

\subsection{Converting the Latitude and Longitude to X and Y Coordinates}
The method for identifying a position on the earth is called the coordinate system. We can use for this case latitude/longitude, easting /northing, and X/Y. The coordinate systems are classified into two categories: geographic and projected.  
A spherical surface is used for the geographic coordinate system and latitude and longitude are used. Our main goal is to plot the Memphis area in 2D, for which we need to convert a sphere into a flat surface. For that, we use the projected coordinate system.
In particular, we use the pyproj python library. 
% And we are successfully able to extract the X and Y coordinates for the latitude and longitude from OSM data for the Memphis area.

\section{Results} 
\label{results}
\begin{figure}[hbt!]
    \centering\includegraphics[width=1\linewidth]{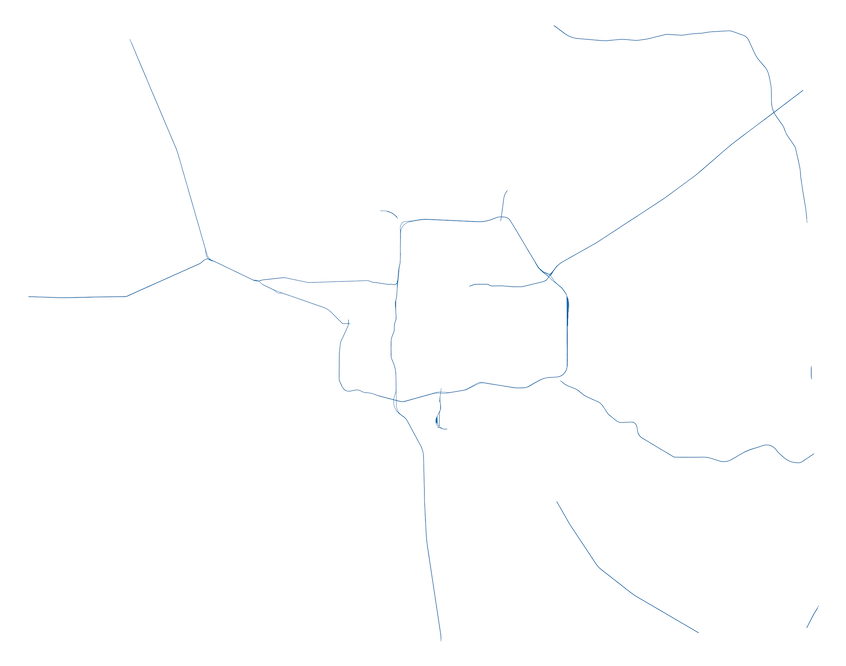}
    \caption{The map showing only the motorway of the OSM Data in Memphis.}
    \label{fig:motorway}
\end{figure}

We conduct a number of experiments to plot the streets of the Memphis area. First, we plot only the motorway and use UTM (Universal Transverse Mercator) zone 15\footnote{\url{https://www.arcgis.com/apps/View/index.html?appid=7fa64a25efd0420896c3336dc2238475}}. 
The plot showing only the motorway can be found in Fig.~\ref{fig:motorway}.

\begin{figure}[hbt!]
    \centering\includegraphics[width=.6\linewidth]{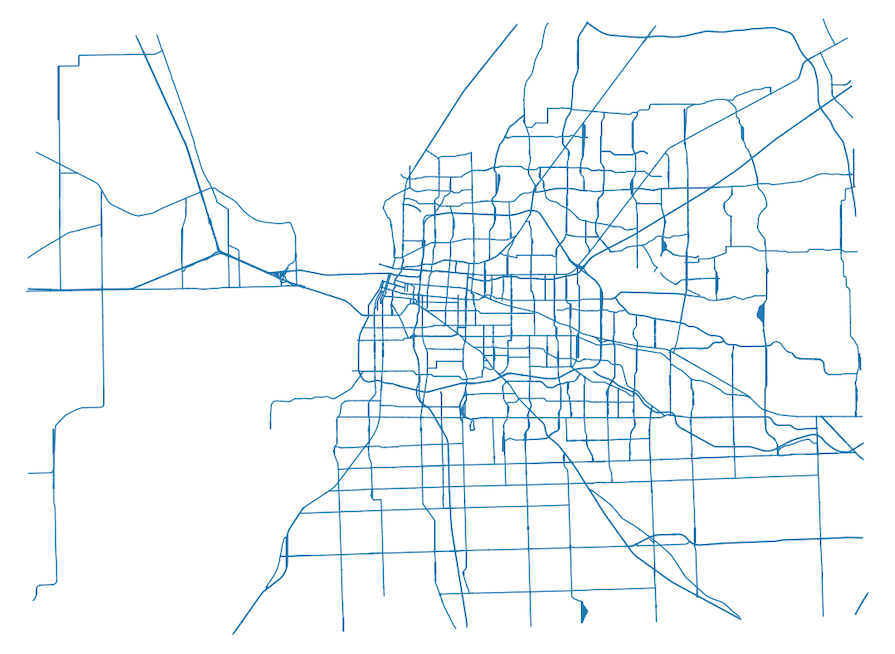}
    \caption{The map showing the motorway, trunk, primary, and secondary ways of Memphis.}
    \label{fig:three streets}
\end{figure}

Next, we plot motorway, trunk, primary, and secondary ways. The plot is shown in Fig.~\ref{fig:three streets}.

\begin{figure}[hbt!]
    \centering\includegraphics[width=.9\linewidth]{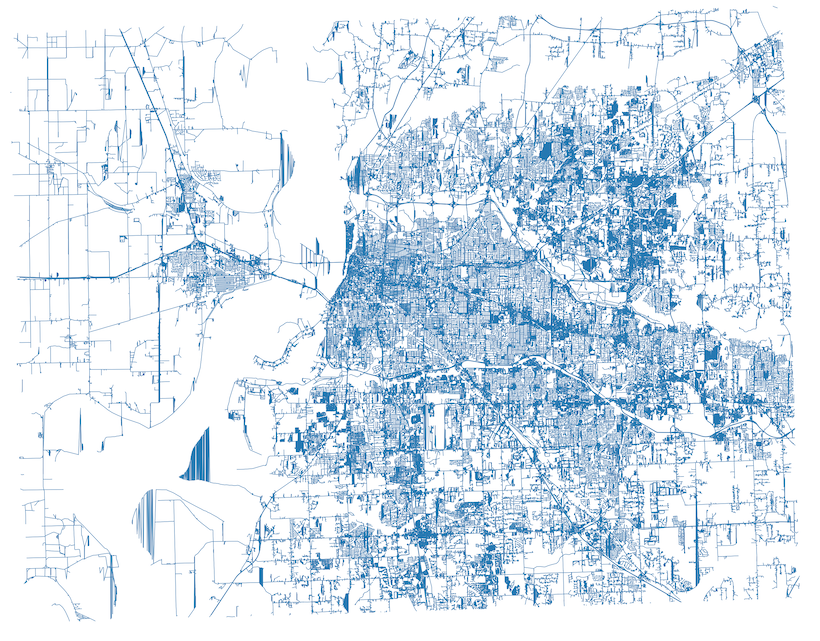}
    \caption{The map containing all ways of Memphis.}
    \label{fig:All}
\end{figure}

Lastly, we plot all ways of Memphis. 
% The runtime for getting the plot for all the streets is enormous because we had huge amounts of X and Y coordinates for all the streets of OSM data. 
The map is shown in Fig.~\ref{fig:All}. 
Note that the map is slightly tilted towards the left-hand side by using the UTM zone 15 which is the UTM zone for Memphis. 
% The Map is slightly tilted towards the left-hand side.

% The observation is that if we are able to plot the map then in the next steps we can apply different Machine Learning (ML) and Nural Networks (NN) for various kinds of applications. 

\section{Conclusion}
In this task, we show how to convert OSM data to road networks which are essential for many downstream mobility applications.
We also visualize Memphis in various ways showing its consisting roads.
% have reported on an analysis of nodes and ways in OSM for getting all the streets for the Memphis area. Then we plotted all the streets for visualizing the maps for all the streets in Memphis city. Nodes and ways are two core elements in the OSM data model. In sections \ref{download} and \ref{results}, we have discussed in detail how we can process the nodes and ways information to get a DataFrame. Then we can easily plot all the streets by grabbing the coordinates from the DataFrame. For more exploration, we will apply different applications for Machine Learning (ML) and Neural Networks (NN). It will give us a huge opportunity to find new dimensions to go and help in different aspects of developing the target city.

% \bibliographystyle{abbrv}
% \bibliography{references} 

\printbibliography

\end{document}